%% file: Lattice2017_263_LEE.tex
\documentclass[epj]{webofc}
\usepackage[utf8]{inputenc}
\usepackage[varg]{txfonts}   
\usepackage{booktabs}
\usepackage{xcolor}
\definecolor{darkred}{rgb}{0.4,0.0,0.0}
\definecolor{darkgreen}{rgb}{0.0,0.4,0.0}
\definecolor{darkblue}{rgb}{0.0,0.0,0.4}
\usepackage[bookmarks,linktocpage,colorlinks,
    linkcolor = darkred,
    urlcolor  = darkblue,
    citecolor = darkgreen]{hyperref}
\usepackage{textcomp}
\usepackage{slashed}
\usepackage[FIGBOTCAP,TABBOTCAP,tight]{subfigure}
\usepackage{comment}
\graphicspath{{./figs/}}
\wocname{EPJ Web of Conferences}
\woctitle{Lattice2017}
\input{macro}
\begin{document}
%
\selectlanguage{english}
\title{
  Update on $\varepsilon_K$ with lattice QCD inputs
}
\author{
\firstname{Yong-Chull} \lastname{Jang}\inst{2}\inst{3} \and
\firstname{Weonjong} \lastname{Lee}\inst{1}\fnsep\thanks{Speaker,
  \email{wlee@snu.ac.kr}} \and
\firstname{Sunkyu} \lastname{Lee}\inst{1} \and
\firstname{Jaehoon} \lastname{Leem}\inst{1} \and
\lastname{(LANL-SWME Collaboration)}
}
\institute{%
  Lattice Gauge Theory Research Center, CTP,
  Department of Physics and Astronomy, \\
  Seoul National University, Seoul 08826, South Korea
  \and
  Los Alamos National Laboratory,
  Theoretical Division T-2, 
  Los Alamos, New Mexico 87545, USA
  \and
  Brookhaven National Laboratory,
  Department of Physics, Upton, New York 11973, USA
}
\abstract{ We report updated results for $\epsK$, the indirect CP
  violation parameter in neutral kaons, which is evaluated directly
  from the standard model with lattice QCD inputs.  We use lattice QCD
  inputs to fix $\BK$, $\abs{\Vcb}$, $\xi_0$, $\xi_2$, $\abs{\Vus}$,
  and $m_c(m_c)$.  Since Lattice 2016, the UTfit group has updated the
  Wolfenstein parameters in the angle-only-fit method, and the HFLAV
  group has also updated $\abs{\Vcb}$.  Our results show that the
  evaluation of $\epsK$ with exclusive $\abs{\Vcb}$ (lattice QCD
  inputs) has $4.0\sigma$ tension with the experimental value, while
  that with inclusive $\abs{\Vcb}$ (heavy quark expansion based on OPE
  and QCD sum rules) shows no tension.  }
\maketitle
\section{Introduction}
\label{sec:intro}

This paper is an update of our previous papers \cite{ Bailey:2015tba,
  Bailey:2015frw, Lee:2016xkb, Bailey:2016dzk}.
In the standard model, the indirect CP violation parameter $\epsK$ of
neutral kaons can be expressed as follows,
\begin{align}
  \label{eq:epsK_def}
  \epsK
  & \equiv \frac{\mathcal{A}(K_L \to \pi\pi(I=0))}
              {\mathcal{A}(K_S \to \pi\pi(I=0))}
  \nonumber \\
  & =  e^{i\theta} \sqrt{2}\sin{\theta}
  \Big( C_{\eps} \hat{B}_{K} X_\text{SD}
  + \frac{ \xi_{0} }{ \sqrt{2} } + \xi_\text{LD} \Big) 
   + \mathcal{O}(\omega\eps^\prime)
   + \mathcal{O}(\xi_0 \Gamma_2/\Gamma_1) \,, 
\end{align}
where
\begin{align}
  C_{\eps} 
  &= \frac{ G_{F}^{2} F_K^{2} m_{K^{0}} M_{W}^{2} }
  { 6\sqrt{2} \; \pi^{2} \; \Delta M_{K} } \,,
  \qquad
  \xi_0 = \frac{\Im A_0}{\Re A_0} \,,
  \qquad
  \xi_\text{LD} =  \frac{m^\prime_\text{LD}}{\sqrt{2} \; \Delta M_K} \,,
  \label{eq:xiLD}
  \\
  m^\prime_\text{LD}
  &= -\Im \left[ \mathcal{P} \; \sum_{C} 
    \frac{\mate{\wbar{K}^0}{H_\text{w}}{C} \mate{C}{H_\text{w}}{K^0}}
         {m_{K^0}-E_{C}}
         \right]
  \label{eq:mLD}
  \\
  X_\text{SD} &= \Im\lambda_t \Big[ \Re\lambda_c \eta_{cc} S_0(x_c)
    -\Re\lambda_t \eta_{tt} S_0(x_t) - (\Re\lambda_c -
    \Re\lambda_t) \eta_{ct} S_0(x_c,x_t) \Big] \,.
  \label{eq:X_SD}
\end{align}
Here, $X_\text{SD}$ represents the short distance contribution from
the box diagrams.
The formulas for $\lambda_i$, $S_0$, $x_i$, and $\eta_{ij}$ with $i,j =
\{c,t\}$ are given in Ref.~\cite{Bailey:2015tba}.
The $\xi_0$ and $\xi_\text{LD}$ represent the long distance effects
from the absorptive part and the dispersive part, respectively.
Since Lattice 2016, there has been a major update on the Wolfenstein
parameters: $\lambda$, $\bar{\rho}$, and $\bar{\eta}$ from the
angle-only fit (AOF) of the UTfit collaboration as well as those from
the global unitarity triangle (UT) fit of the CKMfitter and UTfit
collaborations.
Since Lattice 2016, the HFLAV group has also updated results of
$\abs{\Vcb}$ and $\abs{\Vub}$.
Hence, it is time to update the current status of $\epsK$ in
lattice QCD.
%

\section{Input parameters}
\label{sec:input-para}

Wolfenstein parameters, $\hat{B}_K$, $\lvert V_{cb} \rvert$, $\xi_0$,
and $\xi_\text{LD}$ are separately discussed in the following
subsections.
Other input parameters in Eq.~\eqref{eq:epsK_def} are the same as in
Ref.~\cite{ Bailey:2015tba} except for the charm quark mass $m_c(m_c)
= 1.2733(76)\;\mathrm{GeV}$, which is taken from the HPQCD
collaboraion result \cite{ Chakraborty:2014aca}.
They are given in Tables \ref{tab:eta} and \ref{tab:other}.

\input{tab_eta}

\input{tab_rest}

\subsection{Wolfenstein parameters}
\label{ssec:wolf-para}
%

\input{tab_wp}
The CKMfitter and UTfit collaboration provide the Wolfenstein parameters
($\lambda$, $\bar{\rho}$, $\bar{\eta}$, $A$) determined by the global
unitarity triangle (UT) fit.
The 2017 results are summarized in Table \ref{tab:wp}. 
Here, $\epsK$, $\BK$, and $|V_{cb}|$ are used as inputs to the
global UT fit.
Hence, the Wolfenstein parameters extracted by the global UT fit
contain unwanted correlation with $\epsK$.
In order to avoid this correlation, we take another input set from the
angle-only fit (AOF) suggested in Ref.~\cite{Bevan2013:npps241.89}.
The AOF does not use $\epsK$, $\BK$, and $|V_{cb}|$ as input to
determine the UT apex ($\bar{\rho}$, $\bar{\eta}$).
We can determine $\lambda$ independently from $|V_{us}|$ which comes from
the $K_{\ell 3}$ and $K_{\ell 2}$ decays using lattice QCD inputs.
The 2017 results of the AOF are summarized in Table \ref{tab:wp}.

\subsection{Input parameter $\BK$}
\label{ssec:bk}

Recently, FLAG has reported results for $\BK$ calculated in lattice QCD
with $N_f=2+1$ flavors \cite{Aoki:2016frl}.
\begin{align}
\BK &= 0.7625(97) \qquad \text{from FLAG-2017.}
\end{align}
This is the global average over the results of BMW-2011 \cite{
  Durr2011:PhysLettB.705.477}, Laiho-2011 \cite{ Laiho:2011np},
RBC-UK-2016 \cite{ Blum:2014tka}, and SWME-2016 \cite{Jang:2015sla}.

\subsection{Input parameter $|\Vcb|$}
\label{ssec:V_cb}

\input{tab_Vcb}

\input{fig_Vcb}

In Table \ref{tab:Vcb}, we summarize updated results for both exclusive
$|\Vcb|$ and inclusive $|\Vcb|$.
Recently HFLAV reported them \cite{Amhis:2016xyh}.
At present, we find that there exists a $4.1\sigma$ tension between
exclusive and inclusive $|\Vcb|$ when we use the combined averages
given in Table \ref{tab:Vcb}.
We use these combined results when we evaluate $\epsK$.

In Fig.~\ref{fig:Vcb}, we present results for $|\Vcb|$ and $|\Vub|$.
The big change is that, as of Lattice 2016, the result for exclusive
$|\Vcb|$ from $\bar{B} \to D \ell \bar{\nu}$ was about one sigma
away from that from $\bar{B} \to D^{*} \ell \bar{\nu}$ (refer to
Ref.~\cite{ Lee:2016xkb, Bailey:2016dzk} for more details), but as of
Lattice 2017, they are on top of each other, as shown in
Fig.~\ref{fig:Vcb}\,\subref{fig:Vcb-2017}.
In 2017, due to the addition of more data from ALEPH, CLEO, OPAL, and
DELPHI to the HFLAV analysis for exclusive $|\Vcb|$, the results from
$\bar{B} \to D^{*} \ell \bar{\nu}$ do not change visibly, but those
from $\bar{B} \to D \ell \bar{\nu}$ shift downward by about
$1.0\sigma$.
For more details, refer to Ref.~\cite{Amhis:2016xyh}.

%
%
Regarding $|\Vub|/|\Vcb|$, in 2016, we used the results of lattice QCD
in Ref.~\cite{Detmold:2015aaa}, but in 2017, we use the results of
HFLAV in Ref.~\cite{Amhis:2016xyh}.
Due to the addition of more data to the HFLAV analysis, the results
shift downward by $\frac{3}{4}\sigma$ in 2017.
%

%
%
Regarding $|\Vub|$, in 2016, we used the results of Ref.~\cite{
  DeTar:2015orc} obtained using a combined fit of the results in
Refs.~\cite{ Lattice:2015tia, Flynn:2015mha} for $B \to \pi \ell \nu$
decay over the data subset from BABAR and BELLE, but in 2017, we use
the results of HFLAV in Ref.~\cite{Amhis:2016xyh} obtained using the
exclusive decay results of $B \to \pi \ell \nu$ from BABAR and BELLE.
Both results are used as input to determine $|\Vcb|$ and $|\Vub|$
from the combined fit with all the exclusive decay results for $B$
mesons and $\Lambda_b$ baryons.
The shift in $|\Vub|$ is downward by about $0.1\sigma$ in 2017.

\subsection{Input parameters $\xi_0$ and $\xi_\text{LD}$}
\label{ssec:xi0}
The absorptive part of long distance effects in $\epsK$ is
parametrized into $\xi_0$.
It is related to $\eps'/\eps$ and $\xi_2$ as follows,
\begin{align}
\xi_0  &= \frac{\Im A_0}{\Re A_0}, \qquad 
\xi_2 = \frac{\Im A_2}{\Re A_2}, \qquad
\Re \left(\frac{\eps'}{\eps} \right) = 
\frac{\omega}{\sqrt{2} |\eps_K|} (\xi_2 - \xi_0) \,.
\label{eq:e'/e:xi0}
\end{align}

There are two independent methods to determine $\xi_0$ in lattice QCD:
one is the indirect method and the other is the direct method.
In the indirect method, we determine $\xi_0$ using
Eq.~\eqref{eq:e'/e:xi0} with lattice QCD input $\xi_2$ and with
experimental results for $\eps'/\eps$, $\epsK$, and $\omega$.
In the direct method, we can determine $\xi_0$ directly using
lattice QCD results for $\Im A_0$ combined with experimental results
for $\Re A_0$.
Here, we prefer the indirect method to the direct method thanks to two
reasons. 
The first reason is that the lattice QCD calculation of $\Im A_0$ is
much noisier than that of $\Im A_2$ due to many disconnected diagrams.
The second reason is that the S-wave phase shift $\delta_0$ of the
$\pi-\pi$ scattering in Ref.~\cite{ Bai:2015nea} is lower by
$3.0\sigma$ than the conventional determination of $\delta_0$ in
Ref.~\cite{ Colangelo:2001df, GarciaMartin:2011cn,
  DescotesGenon:2001tn}, which indicates that there might be some
issues unresolved at present.
For more details on the second reason, refer to Ref.~\cite{
  Lee:2016xkb, Bailey:2016dzk}.
In Table \ref{tab:xi0}, we present results of $\xi_0$ determined
using both indirect and direct methods.
Here, we use the value of $\xi_0$ with the indirect method.

\input{tab_xi0}

The $\xi_\text{LD}$ parameter represents the long distance effect
from the dispersive part.
There has been an on-going attempt to calculate it in lattice QCD
\cite{Christ:2014qwa}.
However, this attempt \cite{Christ:2015phf} is an exploratory study
rather than a high precision measurement at present.
Hence, in this paper we use the rough estimate of $\xi_\text{LD}$ in
Ref.~\cite{ Christ:2014qwa}.
It is given in Table \ref{tab:xi0}.

\section{Results for $\epsK$ with lattice QCD inputs}

\input{fig_epsK}

In Fig.~\ref{fig:epsK}, we present results for $\epsK$ calculated
directly from the standard model with the lattice QCD inputs
described in the previous section.
In Fig.~\ref{fig:epsK}\,\subref{fig:ex-epsK}, the blue curve
represents the theoretical evaluation of $\epsK$ with FLAG $\BK$, AOF
Wolfenstein parameters, and exclusive $|\Vcb|$ which corresponds
to ex-combined in Table \ref{tab:Vcb}\,\subref{tab:ex-Vcb}.
The red curve in Fig.~\ref{fig:epsK} represents the experimental value
of $\epsK$.
In Fig.~\ref{fig:epsK}\,\subref{fig:in-epsK}, the blue curve
represents the same as in Fig.~\ref{fig:epsK}\,\subref{fig:ex-epsK}
except for using the inclusive $|\Vcb|$ which corresponds to
in-combined in Table \ref{tab:Vcb}\,\subref{tab:in-Vcb}.

The updated results for $\epsK$ are, in units of $1.0\times
10^{-3}$,
\begin{align}
  |\epsK| &= 1.58 \pm 0.16 & & \text{for exclusive $|\Vcb|$ (lattice QCD)}
  \\
  |\epsK| &= 2.05 \pm 0.18 & & \text{for inclusive $|\Vcb|$ (QCD sum rule)}
  \\  
  |\epsK| &= 2.228 \pm 0.011 & & \text{(experimental value)}  
\end{align}
This indicates that the theoretical evaluation of $\epsK$ with lattice
QCD inputs (with exclusive $|\Vcb|$) has $4.0\sigma$ tension with the
experimental result, while there is no tension in the inclusive
$|\Vcb|$ channel (heavy quark expansion based on the OPE and QCD
sum rules).
%

\input{tab_err}
In Table \ref{tab:epsK-budget}, we present the error budget for
$\epsK^\text{SM}$.
Here, we find that the uncertainty from $|V_{cb}|$ is dominant in the
error budget, while the errors from $\bar{\eta}$ and $\eta_{ct}$ are
sub-dominant.
Hence, if we are to see a gap $\Delta \epsK$ greater than $5.0\sigma$,
it is essential to reduce the error in $|V_{cb}|$ significantly.
To reduce this error as much as possible, a project to calculate
$\bar{B} \to D^{(\ast)} \ell \bar{\nu}$ form factors using the OK
action is underway.
Note that the OK action is improved up to the $\lambda^3$
order\footnote{$\lambda \approx \dfrac{\Lambda}{2 m_Q} \approx
  \dfrac{1}{8} $ for the charm quark.} in the HQET power counting,
while the original Fermilab action is improved up to the $\lambda^1$
order.
For more details on this issue, refer to Refs.~\cite{ SWPark:2017,
  JHLeem:2017, Bailey:2016wza, Jeong:2016eot}.

\input{fig_epsK_his}

In Fig.~\ref{fig:epsK-his}, we plot the $\Delta \epsK =
\epsK^\text{Exp} - \epsK^\text{SM}$ in units of $\sigma$ (= the total
error in the estimate of $\Delta\epsK$) as the time evolves starting
from 2012.
We started to monitor $\Delta \epsK$ in 2012 when several lattice QCD
results for $\BK$ obtained using different methods became consistent
with each other within one sigma.
In 2012, $\Delta \epsK$ was $2.5\sigma$ but now it is $4.0\sigma$.
To understand the time evolution over the past 5 years, we have
performed an additional analysis on the error and average.

\input{fig_depsK_err_his}

In Fig.~\ref{fig:epsK-err-his}\,\subref{fig:depsK-err-his}, we plot
the chronological evolution of $\Delta\epsK$ and its error
$\sigma_{\Delta\epsK}$.
Here, we find that the value for $\Delta\epsK$ has increased with some
fluctuations by 25\% during the period of 2012---2017, and its error
$\sigma_{\Delta\epsK}$ has decreased monotonically by 24\% in the same
period.
These two effects interfere constructively so as to produce the
$4.0\sigma$ tension in $\Delta\epsK$ in 2017.
The monotonic decrease in the error $\sigma_{\Delta\epsK}$ reflects 
the fact that lattice QCD calculations are becoming more precise, and
the experimental results also are becoming more accurate.
In Fig.~\ref{fig:epsK-err-his}\,\subref{fig:epsK-ratio-his}, we show
the time evolution of $\epsK^\text{SM}$ and $\Delta\epsK$ in units of
$\epsK^\text{Exp}$.
Here, we find that the evaluation of $\epsK^\text{SM}$ with lattice
QCD inputs explains only 71\% of the experimental value, and the gap
of about 29\% in $\Delta\epsK$ cannot be described in the standard
model with lattice QCD inputs.
%


\section{Acknowledgement}
\begin{acknowledgement}
  We would like to express our sincere gratitude to Jon Bailey for
  help with the manuscript.
  We would like to express sincere gratitude to Carleton DeTar, Aida
  El-Khadra, and Andreas Kronfeld for helpful discussion.
  We also would like to express sincere gratitude to Guido Martinelli
  for providing to us most updated results of UTfit.
  The research of W.~Lee is supported by the Creative Research
  Initiatives Program (No.~2017013332) of the NRF grant funded by the
  Korean government (MEST).
  W.~Lee would like to acknowledge the support from the KISTI
  supercomputing center through the strategic support program for the
  supercomputing application research [No.~KSC-2014-G2-002].
  Computations were carried out in part on the DAVID GPU clusters at
  Seoul National University.
%
%
\end{acknowledgement}

\bibliography{refs}

\end{document}

%% file: macro.tex
\providecommand{\wbar}[1]{\overline#1}
\providecommand{\abs}[1]{\lvert#1\rvert}

\providecommand{\mate}[3]{\langle#1\lvert#2\rvert#3\rangle}

\renewcommand{\Re}{\mathrm{Re}}
\renewcommand{\Im}{\mathrm{Im}}

\newcommand{\BK}{\hat{B}_{K}}
\newcommand{\Vcb}{V_{cb}}
\newcommand{\Vub}{V_{ub}}
\newcommand{\Vus}{V_{us}}

\newcommand{\eps}{\varepsilon}

\newcommand{\epsK}{\varepsilon_{K}}



%% file: tab_eta.tex
\begin{table}[h]
  \sidecaption
  \centering
  \renewcommand{\arraystretch}{1.2}
  \resizebox{0.30\textwidth}{!}{
    \begin{tabular}[b]{@{\quad} c @{\qquad}l @{\qquad}c @{\quad}}
      \hline\hline
      Input & Value & Ref. \\ \hline
      $\eta_{cc}$ & $1.72(27)$   & \cite{Bailey:2015tba} 
      \\ \hline
      $\eta_{tt}$ & $0.5765(65)$ & \cite{Buras2008:PhysRevD.78.033005} 
      \\ \hline
      $\eta_{ct}$ & $0.496(47)$  & \cite{Brod2010:prd.82.094026}
      \\ \hline\hline
    \end{tabular}
  } 
  \caption{QCD corrections: $\eta_{cc}$, $\eta_{tt}$, and $\eta_{ct}$.}
  \label{tab:eta}
\end{table}

%% file: tab_rest.tex
\begin{table}[h]
  \caption{Other input parameters.}
  \label{tab:other}
  \renewcommand{\subfigcapskip}{0.55em}
  \subtable[]{
    \renewcommand{\arraystretch}{1.2}
    \resizebox{0.45\textwidth}{!}{
      \begin{tabular}{@{\quad} c @{\qquad} l @{\qquad} c @{\quad}}
        \hline\hline
        Input & Value & Ref. \\ \hline
        $G_{F}$
        & $1.1663787(6) \times 10^{-5}$ GeV$^{-2}$
        &\cite{Agashe2014:ChinPhysC.38.090001} \\ \hline
        $M_{W}$
        & $80.385(15)$ GeV
        &\cite{Agashe2014:ChinPhysC.38.090001} \\ \hline
        $m_{c}(m_{c})$
        & $1.2733(76)$ GeV
        &\cite{Chakraborty:2014aca} \\ \hline
        $m_{t}(m_{t})$
        & $163.3(2.7)$ GeV
        &\cite{Alekhin2012:plb.716.214} \\ \hline\hline
      \end{tabular}
      } 
    \label{tab:other-1}
    } 
  \hfill
  \subtable[]{
    \renewcommand{\arraystretch}{1.2}
    \resizebox{0.40\textwidth}{!}{
      \begin{tabular}{@{\quad} c @{\qquad} l @{\qquad} c @{\quad}}
        \hline\hline
        Input & Value & Ref. \\ \hline
        $\theta$
        & $43.52(5)^{\circ}$
        &\cite{Agashe2014:ChinPhysC.38.090001} \\ \hline
        $m_{K^{0}}$
        & $497.614(24)$ MeV
        &\cite{Agashe2014:ChinPhysC.38.090001} \\ \hline
        $\Delta M_{K}$
        & $3.484(6) \times 10^{-12}$ MeV
        &\cite{Agashe2014:ChinPhysC.38.090001} \\ \hline
        $F_K$
        & $156.2(7)$ MeV
        &\cite{Agashe2014:ChinPhysC.38.090001}
        \\ \hline\hline
      \end{tabular}
    } 
    \label{tab:other-2}
  } 
\end{table}

%% file: tab_wp.tex
\begin{table}[h]
  \caption{Wolfenstein parameters.}
  \label{tab:wp}
  \centering
  \resizebox{0.60\textwidth}{!}{
    \begin{tabular}{c|ccc}
      \hline\hline
      & CKMfitter & UTfit & AOF \cite{Bevan2013:npps241.89} \\ \hline
      $\lambda$
      & $0.22509(29)$/\cite{Charles:2004jd}
      & $0.22497(69)$/\cite{Bona:2006ah}
      & $0.2248(6)$/\cite{Patrignani:2016xqp}
      \\ \hline
      $\bar{\rho}$
      & $0.1598(76)$/\cite{Charles:2004jd}
      & $0.153(13)$/\cite{Bona:2006ah}
      & $0.146(22)$/\cite{Martinelli:2017}
      \\ \hline
      $\bar{\eta}$
      & $0.3499(63)$/\cite{Charles:2004jd}
      & $0.343(11)$/\cite{Bona:2006ah}
      & $0.333(16)$/\cite{Martinelli:2017}
      \\ \hline\hline
      \end{tabular}
    } 
\end{table}

%% file: tab_Vcb.tex
\begin{table}[h]
  \renewcommand{\arraystretch}{1.2}
  \renewcommand{\subfigcapskip}{0.55em}
  \caption{Results for $|\Vcb|$ in units of $1.0\times 10^{-3}$}
  \label{tab:Vcb}
  \subtable[Exclusive $|\Vcb|$]{
    \resizebox{0.48\textwidth}{!}{
      \begin{tabular}{l|l|l}
        \hline\hline
        $B\to D^* \ell \bar{\nu}$
        & $39.05(47)(58)$ & \cite{Amhis:2016xyh} HFLAV 2017 \\ \hline
        $B\to D \ell \bar{\nu}$
        & $39.18(94)(36)$ & \cite{Amhis:2016xyh} HFLAV 2017 \\ \hline
        $|V_{ub}|/|V_{cb}|$
        & $0.080(4)(4)$   & \cite{Amhis:2016xyh} HFLAV 2017 \\ \hline\hline
        ex-combined
        & $39.13(59)$     & \cite{Amhis:2016xyh} HFLAV 2017 \\ \hline\hline
      \end{tabular}
    } 
    \label{tab:ex-Vcb} 
  } 
  \hfill
  \subtable[Inclusive $|\Vcb|$]{
    \resizebox{0.48\textwidth}{!}{
      \begin{tabular}{l|l|l}
        \hline\hline
        kinetic scheme & $42.19(78)$ & \cite{Amhis:2016xyh} HFLAV 2017 
        \\ \hline
        1S scheme      & $41.98(45)$ & \cite{Amhis:2016xyh} HFLAV 2017 
        \\ \hline\hline
        in-combined    & $42.03(39)$ & this paper 
        \\ \hline\hline
      \end{tabular}
    } 
    \label{tab:in-Vcb} 
  } 
\end{table}

%% file: fig_Vcb.tex
\begin{figure}[h]
  \hspace{-10mm}
  \subfigure[Lattice 2016]{
    \includegraphics[width=0.55\textwidth]{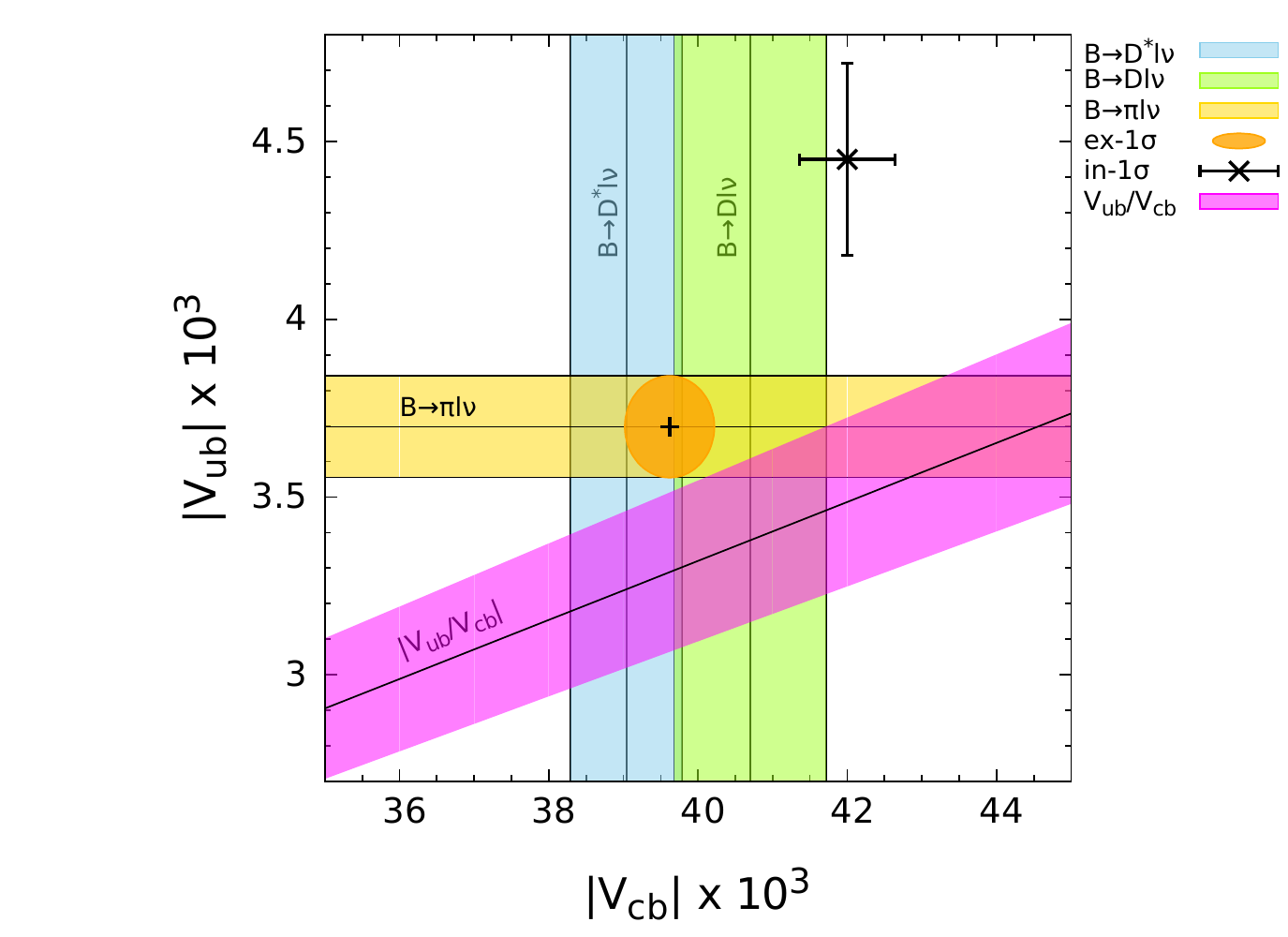}
    \label{fig:Vcb-2016}
  }
  \hfill
  \hspace{-10mm}
  \subfigure[Lattice 2017]{
    \includegraphics[width=0.55\textwidth]{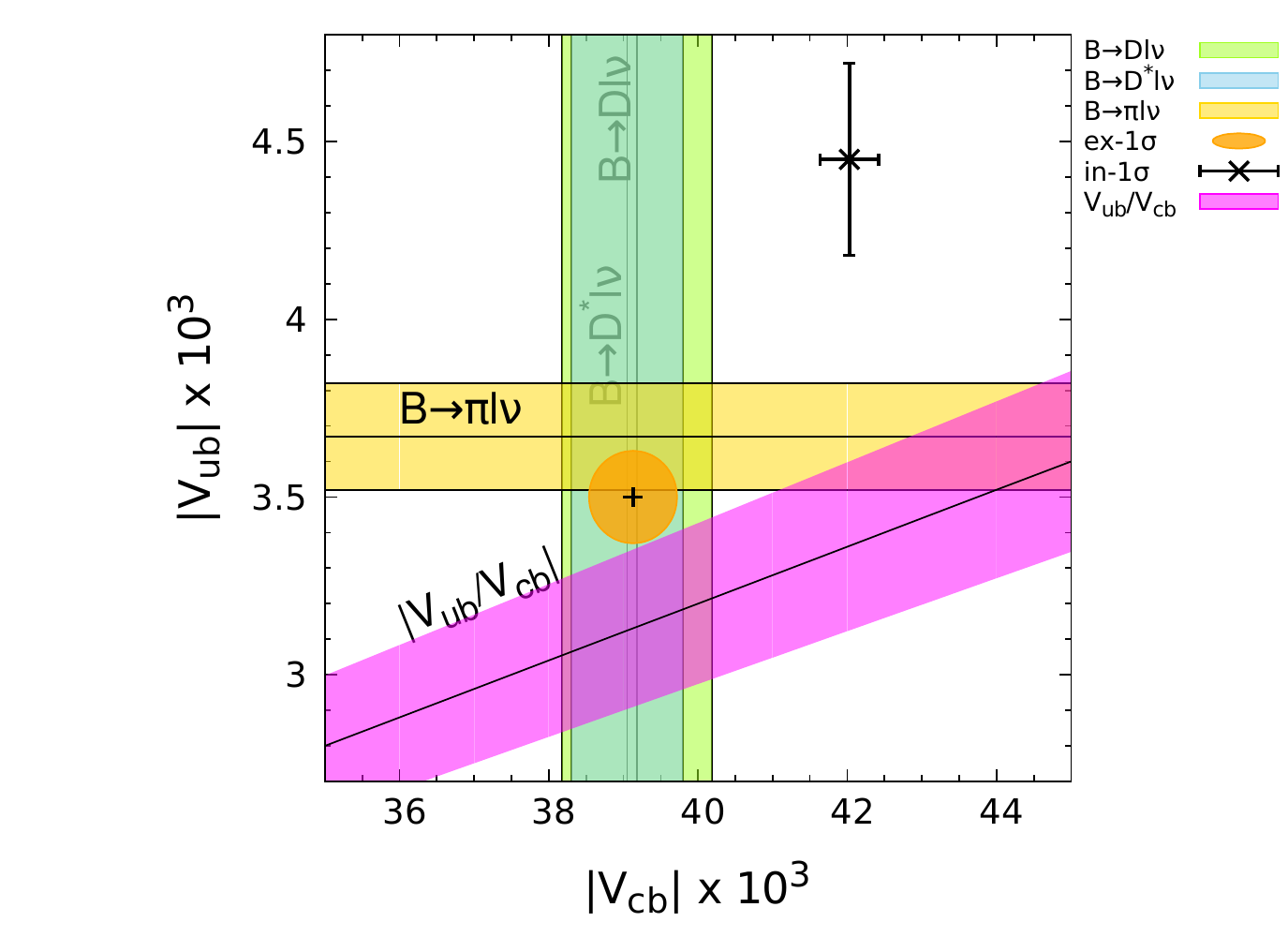}
    \label{fig:Vcb-2017}
  }
  \caption{$|\Vcb|$ versus $|\Vub|$ in units of $1.0\times 10^{-3}$ as
    of \subref{fig:Vcb-2016} Lattice 2016 \cite{ Bailey:2016dzk,
      Lee:2016xkb} and \subref{fig:Vcb-2017} Lattice 2017
    \cite{Amhis:2016xyh}: The light-blue band represents $|V_{cb}|$
    determined from the $\bar{B}\to D^* \ell \bar\nu$ decay mode. The
    light-green band represents $|V_{cb}|$ determined from the
    $\bar{B}\to D\ell\bar\nu$ decay mode. The yellow band represents
    $|V_{ub}|$ determined from the $\bar{B} \to \pi\ell\bar\nu$ decay
    mode.  The magenta band represents $|V_{ub}/V_{cb}|$ determined
    from the LHCb data of the $\Lambda_b \to \Lambda_c \ell\bar\nu$
    and $\Lambda_b \to p \ell\bar\nu$ decay modes. The orange circle
    represents the combined results for exclusive $|V_{cb}|$ and
    $|V_{ub}|$ from the $B$ meson and $\Lambda_b$ decays within
    $1.0\sigma$. The black cross (\textbf{\texttimes}) represents the
    inclusive $|V_{cb}|$ and $|V_{ub}|$ from the heavy quark
    expansion. The details are given in Table \ref{tab:Vcb}.  }
  \label{fig:Vcb}
\end{figure}

%% file: tab_xi0.tex
\begin{table}[h]
  \sidecaption
  \centering
  \renewcommand{\arraystretch}{1.2}
  \resizebox{0.47\textwidth}{!}{
    \begin{tabular}[b]{@{\quad}l @{\qquad} l @{\qquad} l @{\qquad} c @{\quad}}
      \hline\hline
      Input & Method & Value & Ref. \\ \hline
      $\xi_0$ & indirect & $-1.63(19) \times 10^{-4}$
      & \cite{Blum:2015ywa}
      \\ \hline
      $\xi_0$ & direct  & $-0.57(49) \times 10^{-4}$
      & \cite{Bai:2015nea}
      \\ \hline
      $\xi_\text{LD}$ & --- & $(0 \pm 1.6)\,\%$
      & \cite{Christ2012:PhysRevD.88.014508}
      \\ \hline\hline
    \end{tabular}
  } 
  \caption{Long distance effects: $\xi_0$ and $\xi_\text{LD}$. In the
    case of $\xi_\text{LD}$, we do not know the precise value at
    present, and so we quote a rough estimate \cite{Christ:2014qwa}
    incorporated as a systematic error.}
  \label{tab:xi0}
\end{table}

%% file: fig_epsK.tex
\begin{figure}[h]
  \subfigure[$\abs{\epsK}$ with exclusive $|\Vcb|$]{
    \includegraphics[width=0.49\textwidth]{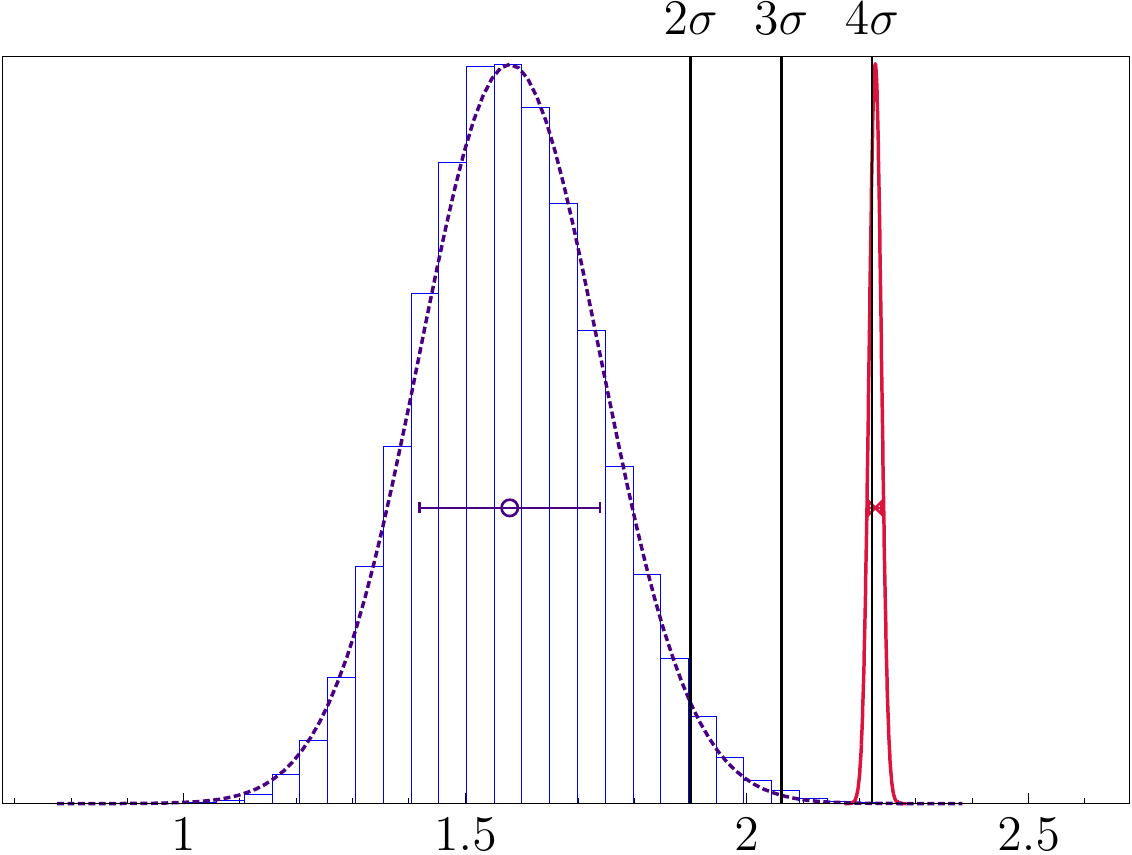}
    \label{fig:ex-epsK}
  }
  \hfill
  \subfigure[$\abs{\epsK}$ with inclusive $|\Vcb|$]{
    \includegraphics[width=0.49\textwidth]{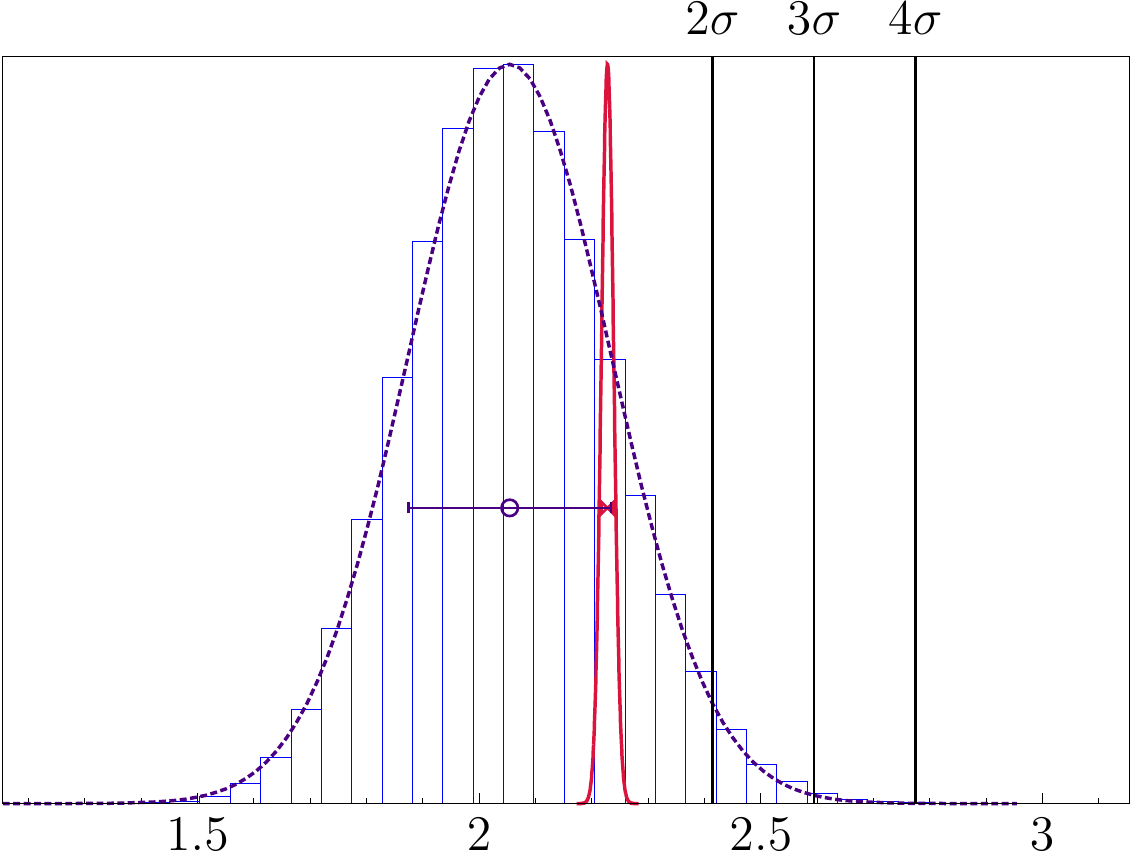}
    \label{fig:in-epsK}
  }
  \caption{$\abs{\epsK}$ with \subref{fig:ex-epsK} exclusive $|\Vcb|$
    (left) and \subref{fig:in-epsK} inclusive $|\Vcb|$ (right) in
    units of $1.0 \times 10^{-3}$. Here, we use the FLAG-2017 $\BK$
    and AOF for the Wolfenstein parameters. The red curve represents
    the experimental value of $\epsK$, and the blue curve represents
    the theoretical value calculated directly from the standard
    model. }
  \label{fig:epsK}
\end{figure}

%% file: tab_err.tex
\begin{table}[tb!]
  \sidecaption
  \renewcommand{\arraystretch}{1.2}
  \resizebox{0.35\textwidth}{!}{
    \begin{tabular}[b]{ccc}
      \hline\hline
      source           & error (\%) & memo \\
      \hline
      $V_{cb}$         & 30.1        & ex-combined \\
      $\bar{\eta}$     & 25.8        & AOF \\
      $\eta_{ct}$      & 20.8        & $c-t$ Box \\
      $\eta_{cc}$      &  8.8        & $c-c$ Box \\
      $\bar{\rho}$     &  3.9        & AOF \\
      $m_t$            &  3.0        & \\
      $\xi_\text{LD}$  &  2.5        & RBC/UKQCD\\
      $\hat{B}_K$      &  1.8        & FLAG-2017 \\
      $\xi_0$          &  1.3        & RBC/UKQCD\\
      $\vdots$         & $\vdots$    & \\
      \hline\hline
    \end{tabular}
  } 
  \caption{ Error budget for $\epsK^\text{SM}$ obtained using the AOF
    method, the exclusive $\abs{\Vcb}$, and the FLAG $\BK$. Here, the
    values in the error column are fractional contributions in units
    of percentage (\%) to the total error obtained using the formula
    given in Ref.~\cite{ Bailey:2015tba}. Hence, the total sum of the
    errors(\%) will be 100\%.}
  \label{tab:epsK-budget}
\end{table}

%% file: fig_epsK_his.tex
\begin{figure}[h]
  \sidecaption
  \includegraphics[width=0.55\textwidth]{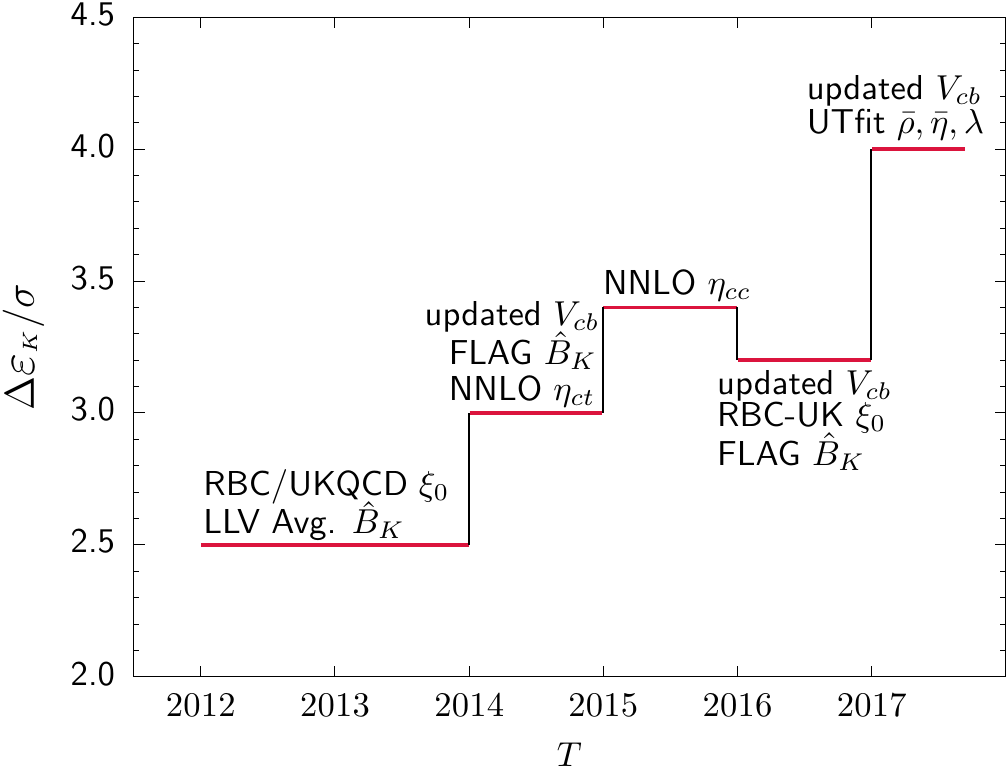}
  \caption{ Chronological evolution of $\Delta \epsK$ in units of
    $\sigma$.  Here, $\Delta \epsK \equiv \epsK^\text{Exp} -
    \epsK^\text{SM}$, where $\epsK^\text{Exp}$ represents the
    experimental value of $\epsK$, and $\epsK^\text{SM}$ represents
    the theoretical evaluation of $\epsK$ calculated directly from the
    standard model with lattice QCD inputs. Here, $1.0\sigma$
    represents the statistical and systematic uncertainty in the
    estimation of $\Delta\epsK$.  }
  \label{fig:epsK-his}
\end{figure}

%% file: fig_depsK_err_his.tex
\begin{figure}[h]
  \subfigure[history of $\Delta \epsK$ and $\sigma_{\Delta\epsK}$]{
    \includegraphics[width=0.48\textwidth]{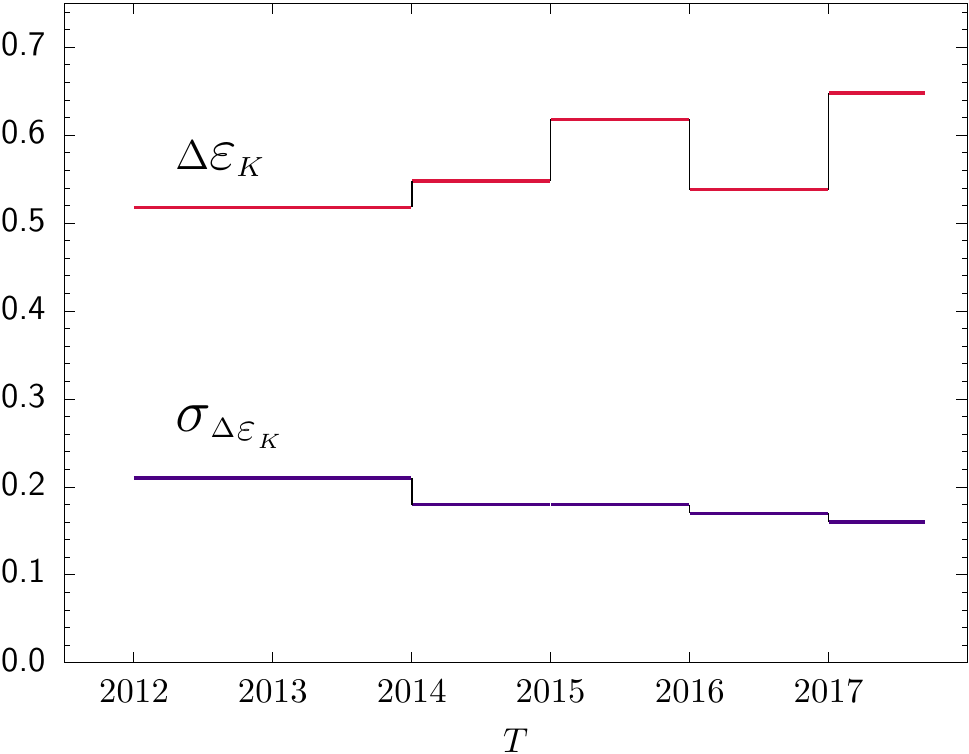}
    \label{fig:depsK-err-his}
  }
  \hfill
  \subfigure[history of $\epsK^\text{SM}$, and $\Delta\epsK$]{
    \includegraphics[width=0.48\textwidth]{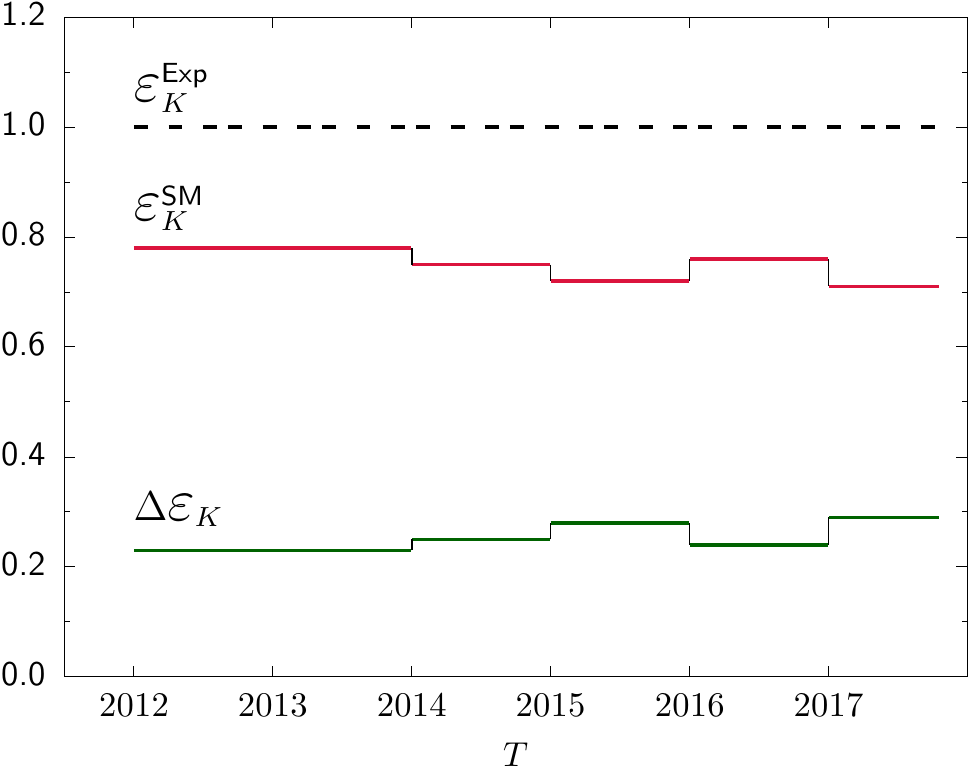}
    \label{fig:epsK-ratio-his}
  }
  \caption{ Chronological evolution of \subref{fig:depsK-err-his}
    $\Delta \epsK$ and $\sigma_{\Delta\epsK}$ in units of $1.0\times
    10^{-3}$, and \subref{fig:epsK-ratio-his} $\epsK^\text{SM}$ and
    $\Delta\epsK$ in units of $\epsK^\text{Exp}$.  Here, $\Delta \epsK
    \equiv \epsK^\text{Exp} - \epsK^\text{SM}$, where
    $\epsK^\text{Exp}$ represents the experimental value of $\epsK$,
    and $\epsK^\text{SM}$ represents the theoretical evaluation of
    $\epsK$ calculated directly from the standard model with lattice
    QCD inputs. Here, $\sigma_{\Delta\epsK}$ represents the
    statistical and systematic uncertainty in the estimation of
    $\Delta\epsK$.  }
  \label{fig:epsK-err-his}
\end{figure}